# Transport Spin Polarization of Noncollinear Antiferromagnetic Antiperovskites

Gautam Gurung, Ding-Fu Shao,* and Evgeny Y. Tsymbal†

*Department of Physics and Astronomy & Nebraska Center for Materials and Nanoscience,*
*University of Nebraska, Lincoln, Nebraska 68588-0299, USA*

Spin-polarized currents play a key role in spintronics. Recently, it has been found that antiferromagnets with a non-spin-degenerate band structure can efficiently spin-polarize electric currents, even though their net magnetization is zero. Among the antiferromagnetic metals with magnetic space group symmetry supporting this functionality, the noncollinear antiferromagnetic antiperovskites $A$NMn$_3$ ($A$ = Ga, Ni, Sn, and Pt) are especially promising. This is due to their high Néel temperatures and a good lattice match to perovskite oxide substrates, offering possibilities of high structural quality heterostructures based on these materials. Here, we investigate the spin polarization of antiferromagnetic $A$NMn$_3$ metals using first-principles density functional theory calculations. We find that the spin polarization of the longitudinal currents in these materials is comparable to that in widely used ferromagnetic metals, and thus can be exploited in magnetic tunnel junctions and spin transfer torque devices. Moreover, for certain film growth directions, the out-of-plane transverse spin currents with a giant charge-to-spin conversion efficiency can be achieved, implying that the $A$NMn$_3$ antiperovskites can be used as efficient spin sources. These properties make $A$NMn$_3$ compounds promising for application in spintronics.

Spintronics exploits the spin degree of freedom in electronic devices for information processing and storage [1]. The magnetic order parameter is used as the state variable in these devices, and its detection and manipulation manifest the read and write operations of the stored information. Currents with sizable spin polarization play a central role in the electric performance of such operations used in realistic nanoscale spintronic devices. For example, in magnetic tunnel junctions (MTJs), which are employed in commercial magnetic random-access memories (MRAMs) [2], the electrical detection is realized via the tunneling magnetoresistance (TMR) effect signifying a response of the longitudinal spin-polarized charge current to the relative magnetization orientation of the two ferromagnetic electrodes [3-5]. On the other hand, the electric manipulation of magnetization can be achieved via a spin transfer torque driven by a longitudinal spin-polarized charge current [6,7] or via a spin Hall effect [8-10] where a transverse pure spin current is generated by spin-orbit coupling.

Ferromagnetic metals have been widely used in spintronics due to their finite magnetization which can easily spin-polarize electric currents. More recently, it was argued that antiferromagnetic spintronics is more promising, due to antiferromagnets being robust against magnetic perturbations, producing no stray fields, and exhibiting ultrafast spin dynamics [11-13]. Nevertheless, until recently antiferromagnets have been rarely considered efficient to generate spin-polarized currents. This is because most antiferromagnets exhibit a combined $\hat{T}\hat{O}$ symmetry, where $\hat{T}$ is the time reversal symmetry and $\hat{O}$ is a crystal symmetry. The $\hat{T}\hat{O}$ symmetry enforces Kramers' spin degeneracy and hence vanishing magnetization. While the antiferromagnetic order may lower the symmetry to support some unconventional spin Hall current useful for spin-orbit torque devices [14], the efficiency of the intrinsic charge-to-spin conversion of antiferromagnets [14-16] do not show obvious advantages compared to those of the widely used nonmagnetic heavy metal spin sources [15-17].

Recently, it was found that the $\hat{T}\hat{O}$ symmetry in antiferromagnets can be broken by the noncollinear magnetic order [18] or non-centrosymmetric arrangement of nonmagnetic atoms [19]. The broken $\hat{T}\hat{O}$ symmetry was shown to result in interesting electronic, magnetic and transport properties that previously were only known for ferromagnets, such as the anomalous Hall effect [18-34], the non-relativistic Zeeman-like band splitting [35-37], and the unconventional charge-to-spin conversion [38-43]. The emergence of electric currents with sizable spin polarization is particularly exciting, due to the possible use of these currents and the antiferromagnets generating them in spintronic devices [44-46].

Among the antiferromagnetic material candidates exhibiting the required magnetic space group symmetry to produce spin-polarized currents, Mn-based antiperovskite nitrides $A$NMn$_3$ ($A$ = Ga, Ni, Sn, and Pt) [47] have a few advantages. In particular, these antiferromagnetic metals have sufficiently high Néel temperatures, often about room temperature [47,48]. Also their crystal structures match well to those of the widely used perovskite oxides, which allows the realization of high-quality epitaxial heterostructures for device fabrication [14,49].

In this work, based on first-principles density functional theory calculations, we explore the spin polarization of $A$NMn$_3$ ($A$ = Ga, Ni, Sn, and Pt) antiferromagnetic metals [50]. We find that the longitudinal charge currents passing through $A$NMn$_3$ can have a sizable spin polarization, which allows using antiperovskites in magnetic tunnel junctions and spin transfer torque devices. Moreover, we show that the out-of-plane transverse spin currents with giant charge-to-spin conversion efficiencies can be achieved by controlling the film growth direction. These properties make $A$NMn$_3$ compounds promising for application in spintronics.

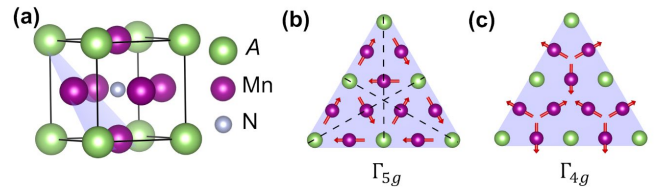

**FIG. 1.** (a) Atomic structure of antiperovskite $A$NMn$_3$. (b, c) The noncollinear alignment of Mn moments in the (111) plane for the antiferromagnteic $\Gamma_{5g}$ (b) and $\Gamma_{4g}$ (c) phases. Red arrows denote magnetic moments and dashed black lines indicate mirror planes $\hat{M}_{0\bar{1}1}$, $\hat{M}_{10\bar{1}}$, and $\hat{M}_{\bar{1}10}$ for the $\Gamma_{5g}$ phase.

**Table 1.** $\hat{T}$-odd spin conductivity tensors for the $\Gamma_{5g}$ and $\Gamma_{4g}$ phases of $A$NMn$_3$. The $x$, $y$, and $z$ axes are set along the [100], [010] and [001] directions for a (001)-stacked film and along the [$\bar{1}10$], [001], and [110] directions for a (110)-stacked film. The matrix elements in the parentheses only appear in the presence of spin-orbit coupling.

| | | $\boldsymbol{\sigma}^x = \begin{bmatrix} \sigma^x_{xx} & \sigma^x_{xy} & \sigma^x_{xz} \\ \sigma^x_{yx} & \sigma^x_{yy} & \sigma^x_{yz} \\ \sigma^x_{zx} & \sigma^x_{zy} & \sigma^x_{zz} \end{bmatrix}$ | $\boldsymbol{\sigma}^y = \begin{bmatrix} \sigma^y_{xx} & \sigma^y_{xy} & \sigma^y_{xz} \\ \sigma^y_{yx} & \sigma^y_{yy} & \sigma^y_{yz} \\ \sigma^y_{zx} & \sigma^y_{zy} & \sigma^y_{zz} \end{bmatrix}$ | $\boldsymbol{\sigma}^z = \begin{bmatrix} \sigma^z_{xx} & \sigma^z_{xy} & \sigma^z_{xz} \\ \sigma^z_{yx} & \sigma^z_{yy} & \sigma^z_{yz} \\ \sigma^z_{zx} & \sigma^z_{zy} & \sigma^z_{zz} \end{bmatrix}$ |
|---|---|---|---|---|
| $A$NMn$_3$ (001) | $\Gamma_{5g}$ | $\begin{bmatrix} 0 & (-a) & (a) \\ (b) & -c & (-d) \\ (-b) & (d) & c \end{bmatrix}$ | $\begin{bmatrix} c & (-b) & (d) \\ (a) & 0 & (-a) \\ (-d) & (b) & -c \end{bmatrix}$ | $\begin{bmatrix} -c & (-d) & (b) \\ (d) & c & (-b) \\ (-a) & (a) & 0 \end{bmatrix}$ |
| | $\Gamma_{4g}$ | $\begin{bmatrix} E & (A) & (A) \\ (B) & C & (D) \\ (B) & (D) & C \end{bmatrix}$ | $\begin{bmatrix} C & (B) & (D) \\ (A) & E & (A) \\ (D) & (B) & C \end{bmatrix}$ | $\begin{bmatrix} C & (D) & (B) \\ (D) & C & (B) \\ (A) & (A) & E \end{bmatrix}$ |
| $A$NMn$_3$ (110) | $\Gamma_{5g}$ | $\begin{bmatrix} \frac{-a+b+c}{\sqrt{2}} & 0 & 0 \\ 0 & -\sqrt{2}c & (b-d) \\ 0 & (-a+d) & \frac{a-b+c}{\sqrt{2}} \end{bmatrix}$ | $\begin{bmatrix} 0 & (-\sqrt{2}b) & c+d \\ (\sqrt{2}a) & 0 & 0 \\ c-d & 0 & 0 \end{bmatrix}$ | $\begin{bmatrix} 0 & (-a-d) & \frac{a+b-c}{\sqrt{2}} \\ (b+d) & 0 & 0 \\ \frac{-a+b+c}{\sqrt{2}} & 0 & 0 \end{bmatrix}$ |
| | $\Gamma_{4g}$ | $\begin{bmatrix} 0 & (A-D) & \frac{A-B-C+E}{\sqrt{2}} \\ (B-D) & 0 & 0 \\ \frac{-A+B-C+E}{\sqrt{2}} & 0 & 0 \end{bmatrix}$ | $\begin{bmatrix} C-D & 0 & 0 \\ 0 & E & (\sqrt{2}A) \\ 0 & (\sqrt{2}B) & C+D \end{bmatrix}$ | $\begin{bmatrix} \frac{-A-B+C+E}{\sqrt{2}} & 0 & 0 \\ 0 & \sqrt{2}C & (B+D) \\ 0 & (A+D) & \frac{A+B+C+E}{\sqrt{2}} \end{bmatrix}$ |

Figure 1(a) shows the crystal structure of the antiperovskite $A$NMn$_3$, which is similar to that of a perovskite except the cation and the anion having swapped positions (Fig. 1(a)). This structure hosts a frustrated Kagomé lattice in the (111) plane that favors noncollinear alignment of the antiferromagnetically coupled magnetic moments [47]. The antiperovskites exhibit interesting properties, such as magnetovolume [66], barocaloric [67], piezomagnetic [62, 63, 68], magnetoelectric [64, 69], anomalous Hall [24 - 29], and unconventional spin Hall [14] effects. One of the most common noncollinear antiferromagnetic orders in $A$NMn$_3$ is $\Gamma_{5g}$ (typical for GaNMn$_3$), where the Mn magnetic moments form a chiral configuration with the 120° angle between each other within the (111) plane (Fig. 1(b)) [47, 61]. $\Gamma_{5g}$ is a compensated antiferromagnetic phase due to three mirror planes $\widehat{M}_{0\bar{1}1}$, $\widehat{M}_{10\bar{1}}$, or $\widehat{M}_{\bar{1}10}$ perpendicular to the (111) plane in the magnetic space group $R\bar{3}m$ which prohibit the net magnetization. Another common noncollinear antiferromagnetic phase is $\Gamma_{4g}$, which can be obtained from the $\Gamma_{5g}$ phase by rotating all magnetic moments about the [111] axis by 90° (Fig. 1(c)) [47]. The mirror symmetries are broken in the $\Gamma_{4g}$ phase so that the corresponding magnetic space group $R\bar{3}m'$ allows an uncompensated magnetization (though very small) and the anomalous Hall effect [24,70].

We find that the magnetic group symmetries of both $\Gamma_{5g}$ and $\Gamma_{4g}$ phases support sizable longitudinal and transverse spin currents. In the diffusive transport regime, the spin conductivity has two contributions: [8,38]

$$\sigma_{ij}^k = -\frac{e\hbar}{\pi}\int \frac{d^3\vec{k}}{(2\pi)^3}\sum_{n,m}\frac{\Gamma^2\text{Re}(\langle n\vec{k}|J_i^k|m\vec{k}\rangle\langle m\vec{k}|v_j|n\vec{k}\rangle)}{\left[(E_F-E_{n\vec{k}})^2+\Gamma^2\right]\left[(E_F-E_{m\vec{k}})^2+\Gamma^2\right]}, \quad (1)$$

and

$$\sigma_{ij}^k = -\frac{2e}{\hbar}\int \frac{d^3\vec{k}}{(2\pi)^3}\sum_{m\neq n}\frac{\text{Im}(\langle n\vec{k}|J_i^k|m\vec{k}\rangle\langle m\vec{k}|v_j|n\vec{k}\rangle)}{(E_{n\vec{k}}-E_{m\vec{k}})^2}. \quad (2)$$

Here $J_i^k = \frac{1}{2}\{v_i, s_k\}$ is the spin current operator, $\Gamma$ is the scattering rate in a constant relaxation time approximation, $f_{n\vec{k}}$ is the Fermi-Dirac distribution function for band $n$ and wave vector $\vec{k}$, $v_i$ and $s_k$ are velocity and spin operators, respectively, and $i$, $j$ and $k$ are the spin-current, charge-current, and spin polarization directions, respectively. The spin conductivity $\sigma_{ij}^k$ given by Eq. (1) is the Fermi surface property odd under time reversal symmetry ($\hat{T}$-odd). As a result, this contribution is allowed only for ferromagnetic and some antiferromagnetic metals without $\hat{T}$ or $\hat{T}\hat{O}$ symmetries, such as noncollinear antiferromagnetic $A$NMn$_3$. In these materials, the Fermi surface is intrinsically spin textured resulting in spin-polarized currents even in the absence of spin-orbit coupling. This leads to finite non-relativistic components of the $\hat{T}$-odd spin conductivity tensor. Spin-orbit coupling alters the spin texture and hence the spin conductivity tensor. However, due to the electronic structure of $A$NMn$_3$ at the Fermi energy ($E_F$) being majorly controlled by the Mn atoms which do not produce strong spin-orbit coupling, these changes are not expected to be significant. (Note though that the spin-orbit coupling makes some $\sigma_{ij}^k$ components finite, which were zero in its absence).

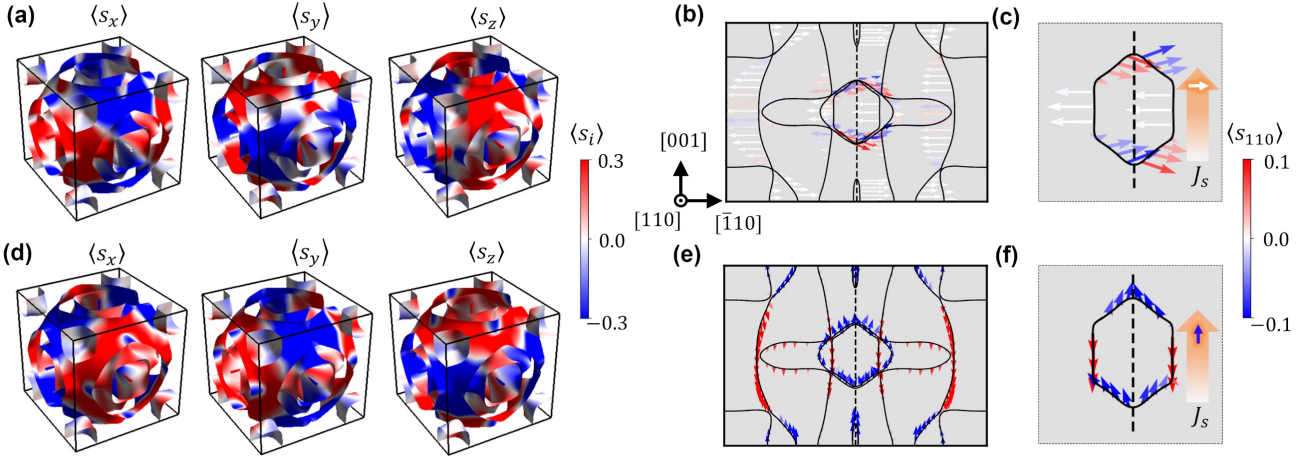

FIG. 2. (a) The spin-projected Fermi surface of GaNMn$_3$ in the $\Gamma_{5g}$ phase. (b) The (110) plane cut of the Fermi surfaces shown in (a), where the solid lines denote the Fermi surface, the colored arrows denote the spin textures, and the dashed line denote the $M_{\bar{1}10}$ mirror plane. (c) The zoomed plot of (b) showing only the central Fermi pocket. (d,e,f) The same in (a,b,c) for GaNMn$_3$ in the $\Gamma_{4g}$ phase.

In contrast to the $\hat{T}$-odd $\sigma_{ij}^k$, the spin conductivity tensor given by Eq. (2) is determined by the interband contributions that are even under time reversal symmetry ($\hat{T}$-even). As a result, non-vanishing $\hat{T}$-even $\sigma_{ij}^k$ components can only appear in the presence of spin-orbit coupling. Therefore, these relativistic components are expected to be small compared to the non-relativistic components of the $\hat{T}$-odd $\sigma_{ij}^k$ (see Supplemental Material [50]). Thus, in the following, we focus only on the $\hat{T}$-odd spin conductivity of $A$NMn$_3$.

Table I displays the $\hat{T}$-odd $\sigma_{ij}^k$ of $A$NMn$_3$ according to the magnetic space group symmetry of the crystal. For a (001)-stacked $A$NMn$_3$ with the $x$, $y$, and $z$ axes aligned along the [100], [010], and [001] crystal directions, there are four independent tensor components (denoted by $a$, $b$, $c$, and $d$) in the $\Gamma_{5g}$ phase. These include the off-diagonal tensor components $\sigma_{ij}^k$ ($i \neq j$) which determine the transverse spin current generated by a longitudinal charge current and are related to the magnetic spin Hall effect proposed recently [38-41]. However, these off-diagonal components are relativistic in nature and are non-zero only in the presence of spin-orbit coupling. Therefore, these relativistic contributions to spin conductivity are not expected to be large, due the weak spin-orbit coupling in $A$NMn$_3$.

There are also non-relativistic diagonal components of the spin conductivity tensor $\sigma_{jj}^k$ ($k \neq j$). These components determine a longitudinal spin current carrying a transverse spin polarization generated by a longitudinal charge current. These diagonal components are non-zero even in the absence of spin-orbit coupling and hence are expected to be large. This implies that a longitudinal spin-polarized charge current can be produced in the compensated antiferromagnetic $\Gamma_{5g}$ phase.

In order to obtain a more intuitive understanding of why the longitudinal charge current is spin-polarized, we use GaNMn$_3$ as a representative example and explore its momentum-dependent spin texture. Due to the broken $\hat{T}$ symmetry in antiferromagnetic GaNMn$_3$, the spin degeneracy is lifted and hence the spin expectation values $\langle s \rangle$ are finite. Since the $\hat{T}$-odd spin conductivity is a purely Fermi surface property (see Eq. (1)), it is the spin texture at the Fermi surface that matters. Figure 2(a) shows the calculated expectation values of the $x$, $y$, and $z$ components of the spin at the Fermi surface of GaNMn$_3$ in the $\Gamma_{5g}$ phase, indicating a rather intricate distribution. The same spin texture, but within the (110) plane, is displayed in Figure 2(b) and enlarged in Figure 2(c) to focus on the Fermi pocket at the center of the Brillouin zone. These spin textures can make the electric currents flowing along certain crystallographic directions of GaNMn$_3$ spin-polarized.

To demonstrate this, we consider symmetry transformations of the spin texture within this Fermi pocket. The inversion-symmetric fragments of the Fermi pocket have the same spin expectation values, since the inversion symmetry $\hat{P}$ does not change the spin:

$$\hat{P}(k_{110}, k_{\bar{1}10}, k_{001}) = (-k_{110}, -k_{\bar{1}10}, -k_{001}),$$
$$\hat{P}(\langle s_{110}\rangle, \langle s_{\bar{1}10}\rangle, \langle s_{001}\rangle) = (\langle s_{110}\rangle, \langle s_{\bar{1}10}\rangle, \langle s_{001}\rangle). \quad (3)$$

Further, the spin expectation values for the wave vectors being symmetric with respect to the $\hat{M}_{\bar{1}10}$ mirror plane have the same $\langle s_{\bar{1}10}\rangle$ components but opposite $\langle s_{001}\rangle$ and $\langle s_{110}\rangle$ components. This is due to the mirror symmetry operation conserving the spin component normal to the mirror plane but flipping the spin component parallel to it:

$$\hat{M}_{\bar{1}10}(k_{110}, k_{\bar{1}10}, k_{001}) = (k_{110}, -k_{\bar{1}10}, k_{001}),$$
$$\hat{M}_{\bar{1}10}(\langle s_{110}\rangle, \langle s_{\bar{1}10}\rangle, \langle s_{001}\rangle) = (-\langle s_{110}\rangle, \langle s_{\bar{1}10}\rangle, -\langle s_{001}\rangle). \quad (4)$$

**Table 2.** Charge-to-spin conversion efficiency $\Phi(\sigma_{ij}^k) = \frac{2e}{\hbar}\sigma_{ij}^k/\sigma_{jj}$ (in %) in antiperovskite $A$NMn$_3$ ($A$ = Ga, Ni, Sn, Pt) in $\Gamma_{5g}$ and $\Gamma_{4g}$ phases for (001)- and (110)-stacked films. Calcuationas are preformed in the presence of spin-orbit coupling for $\Gamma = 0.05$ eV. The $x$, $y$, and $z$ are set along [100], [010] and [001] directions for (001)-stacked film and along [$\bar{1}$10], [001], and [110] directions for a (110)-stacked film.

| | | $\Gamma_{5g}$ | | | | $\Gamma_{4g}$ | | | |
|---|---|---|---|---|---|---|---|---|---|
| | | GaNMn$_3$ | NiNMn$_3$ | SnNMn$_3$ | PtNMn$_3$ | GaNMn$_3$ | NiNMn$_3$ | SnNMn$_3$ | PtNMn$_3$ |
| $A$NMn$_3$ (001) | $\Phi(\sigma_{zz}^x)$ | −20.7 | 13.3 | −27.0 | 14.2 | −12.3 | 7.7 | −16.0 | 10.2 |
| | $\Phi(\sigma_{zz}^z)$ | 0.0 | 0.0 | 0.0 | 0.0 | 24.4 | −15.1 | 31.2 | −18.1 |
| | $\Phi(\sigma_{zx}^x)$ | −0.4 | −0.7 | −0.7 | −3.4 | −0.1 | −0.3 | 0.6 | −1.8 |
| | $\Phi(\sigma_{zx}^y)$ | −0.3 | −0.4 | −0.6 | −0.01 | −0.8 | −0.3 | −0.4 | −3.8 |
| | $\Phi(\sigma_{zx}^z)$ | 0.1 | 0.5 | −0.1 | 1.0 | −0.3 | −0.2 | 0.3 | −0.1 |
| $A$NMn$_3$ (110) | $\Phi(\sigma_{zz}^x)$ | −15.0 | 8.6 | −19.5 | 6.9 | 0.0 | 0.0 | 0.0 | 0.0 |
| | $\Phi(\sigma_{zz}^z)$ | 0.0 | 0.0 | 0.0 | 0.0 | 8.3 | −5.6 | 11.4 | −6.9 |
| | $\Phi(\sigma_{zx}^x)$ | 0.0 | 0.0 | 0.0 | 0.0 | 26.1 | −16.2 | 33.6 | −21.2 |
| | $\Phi(\sigma_{zx}^y)$ | −21.0 | 12.9 | −27.6 | 14.2 | 0.0 | 0.0 | 0.0 | 0.0 |
| | $\Phi(\sigma_{zx}^z)$ | 14.4 | −9.5 | 18.5 | −11.7 | 0.0 | 0.0 | 0.0 | 0.0 |

As a result, the longitudinal electric current parallel to the ($\bar{1}$10) plane, such as the current along the [001] direction shown in Fig. 2(c) is polarized by this spin texture. The associated spin current $J_c$ has a finite $\langle s_{\bar{1}10}\rangle$ component but zero $\langle s_{110}\rangle$ and $\langle s_{001}\rangle$ components, since only $\langle s_{\bar{1}10}\rangle$ is even with respect to $\hat{M}_{\bar{1}10}$. This implies finite matrix elements of the longitudinal spin conductivity tensor $\sigma_{zz}^x = -\sigma_{zz}^y = c$, as shown in Table I for $A$NMn$_3$ in the $\Gamma_{5g}$ phase.

In contrast to the $\Gamma_{5g}$ phase, the $\Gamma_{4g}$ phase of $A$NMn$_3$ has five independent components (denoted by $A$, $B$, $C$, $D$, and $E$) of the $\hat{T}$-odd spin conductivity tensor for a (001)-stacked $A$NMn$_3$. We find that both the diagonal components with spin polarization normal to the charge current direction, $\sigma_{jj}^k = C$ ($k \neq j$), and those parallel to it, $\sigma_{jj}^j = E$, do not vanish in the absence of spin-orbit coupling. This can be illustratively understood by analyzing the spin projected Fermi surfaces of the $\Gamma_{4g}$ GaNMn$_3$ (Fig. 2(d-f)). The spin textures in the $\Gamma_{4g}$ phase are very different from those in the $\Gamma_{5g}$ phase due to different magnetic-space group symmetry. The mirror $\hat{M}_{\bar{1}10}$ plane is broken in the $\Gamma_{4g}$ phase, while a combined $\hat{T}\hat{M}_{\bar{1}10}$ symmetry is preserved, which transforms the wave vector and the spin as follows:

$$\hat{T}\hat{M}_{\bar{1}10}(k_{110}, k_{\bar{1}10}, k_{001}) = (-k_{110}, k_{\bar{1}10}, -k_{001}),$$
$$\hat{T}\hat{M}_{\bar{1}10}(\langle s_{110}\rangle, \langle s_{\bar{1}10}\rangle, \langle s_{001}\rangle) = (\langle s_{110}\rangle, -\langle s_{\bar{1}10}\rangle, \langle s_{001}\rangle). \quad (5)$$

This symmetry together with inversion symmetry $\hat{P}$ (Eq. (3)) implies that $\langle s_{\bar{1}10}\rangle$ is antisymmetric and $\langle s_{110}\rangle$ and $\langle s_{001}\rangle$ are symmetric with respect to ($\bar{1}$10) plane. Therefore, a longitudinal electric current parallel to the ($\bar{1}$10) plane, such as that along the [001] direction, becomes spin-polarized with finite $\langle s_{110}\rangle$ and $\langle s_{001}\rangle$ components and a zero $\langle s_{\bar{1}10}\rangle$ component. This implies finite longitudinal spin conductivities $\sigma_{zz}^x = \sigma_{zz}^y = C$ and $\sigma_{zz}^x = \sigma_{zz}^z = E$ as shown in Table I.

The efficiency of the $\hat{T}$-odd spin current generation can be estimated by calculating a percentage spin conductivity ratio $\Phi(\sigma_{ij}^k) = \frac{2e}{\hbar}\sigma_{ij}^k/\sigma_{jj}$. Here $\sigma_{jj}$ is a conductivity of the longitudinal charge current $J_c$ used to generate the spin current $J_s$ with conductivity $\sigma_{ij}^k$, which can be calculated by replacing the spin current operator $J_i^k$ in Eq. (1) by the charge current operator $J_i = -ev_i$. $\Phi(\sigma_{zj}^k)$ serves as a figure of merit for the performance of realistic spintronic devices. In particular, $\Phi(\sigma_{zx}^k)$ represents the spin Hall angle in spin-torque devices with current-in-plane (CIP) geometry, where an out-of-plane spin current is generated by an in-plane charge current. Similarly, $\Phi(\sigma_{zz}^k)$ measures the degree of spin polarization for an out-of-plane charge current in devices with current-perpendicular-to-plane (CPP) geometry, such as MTJs.

We calculate $\Phi(\sigma_{zj}^k)$ for $A$NMn$_3$ ($A$ = Ga, Ni, Sn, Pt) compounds assuming they are stacked in the (001) plane. Table 2 shows the calculated results for $\Gamma = 0.05$ eV which provides realistic conductivity $\sigma_{zz}$ of the compounds. For the longitudinal spin conductivity, we find that $\Phi(\sigma_{zz}^i)$ is sizable for all $A$NMn$_3$ antiferromagnets we investigated. Especially, we obtain $\Phi(\sigma_{zz}^x) = -20.7\%$ for GaNMn$_3$ which exhibits a $\Gamma_{5g}$ ground state, and $\Phi(\sigma_{zz}^x) = -16.0\ \%$ and $\Phi(\sigma_{zz}^z) = 31.2\%$ for SnNMn$_3$ which has a high-temperature $\Gamma_{4g}$ state. These sizable spin polarizations of the longitudinal current in antiferromagnetic antiperovskites are comparable to those in ferromagnetic metals, such as Fe, Co, and Ni [71 - 73], indicating their potential for spintronic applications, such as antiferromagnetic tunnel junctions discussed below. We note that our results are robust with respect to disorder scattering, as follows from our calculations of $\Phi(\sigma_{zz}^k)$ as a function of $\Gamma$ shown in Supplementary Figure S1 [50].

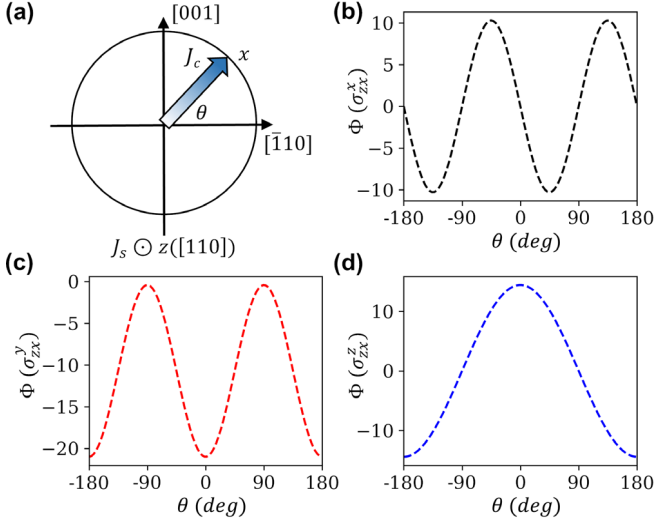

**FIG. 3. (a)** The schematic of charge-to-spin conversion in an $A$NMn$_3$ (110) film. An out-of-plane spin current $J_s$ along the [110] ($z$) direction is generated by applying an in-plane longitudinal charge current $J_c$ along $x$ direction which is away from $[\bar{1}10]$ direction by the angle $\theta$. **(b, c, d)** Charge-to-spin conversion efficiency $\Phi(\sigma_{zx}^k)$ as a function of the longitudinal charge current direction for GaNMn$_3$ (110)-stacked films in the $\Gamma_{5g}$ phase.

On the contrary, the transverse spin conductivity ratio $\Phi(\sigma_{zx}^k)$ is negligible for $A$NMn$_3$ (001) (Table 2). This is understandable since the non-relativistic spin texture does not contribute to $\sigma_{zx}^k$ and the effect appears entirely due to small spin-orbit coupling. However, although $A$NMn$_3$ is not efficient for generating transverse spin currents in (001)-stacked films, it can form a good spin current source for spin-torque devices by engineering the $A$NMn$_3$ growth direction. The related spin conductivity tensors for a film with different orientation can be obtained by applying transformation

$$\sigma_{i'j'}^{k'} = \sum_{k,i,j} R_{k'k} R_{i'i} R_{j'j} \sigma_{ij}^k, \quad (S1)$$

where $\sigma_{ij}^k$ is the spin conductivity for a (001)-stacked film with coordinate system $(x,y,z)$, $\sigma_{i'j'}^{k'}$ is the spin conductivity of a film with coordinate system $(x',y',z')$, $R$ is the rotation matrix to transform the coordinate system $(x,y,z)$ to $(x',y',z')$. A non-relativistic spin texture contribution to $\Phi(\sigma_{zx}^k)$ may be allowed by the magnetic-space group symmetry after the rotation transformation.

As an example, Table 1 shows the spin conductivity tensor for $A$NMn$_3$ (110) in the $\Gamma_{5g}$ and $\Gamma_{4g}$ phases. It is seen that when the charge current direction ($x$) is along the $[\bar{1}10]$ direction, finite $\sigma_{zx}^y$ and $\sigma_{zx}^z$ appear in the $\Gamma_{5g}$ phase, and a finite $\sigma_{zx}^x$ appear in the $\Gamma_{4g}$ phase, even in the absence of spin-orbit coupling. Table 2 shows calculated $\Phi(\sigma_{zx}^k)$ for $A$NMn$_3$ ($A$ = Ga, Ni, Sn, and Pt) (110) films. For the $\Gamma_{5g}$ ground state in GaNMn$_3$, we find large $\Phi(\sigma_{zx}^y) = -21.0\%$ and $\Phi(\sigma_{zx}^z) = 14.4\%$, which are comparable or even larger than these for the reported spin Hall angle in widely used spin source materials such as Pt [74-76].

Moreover, the transverse $\sigma_{zx}^k$ component in the $\Gamma_{5g}$ phase can be engineered by tilting the in-plane longitudinal current direction $x$ from [110] by an angle $\theta$ (Fig. 3(a)). In this case, the $\sigma_{zx}^k$ components are functions of $\theta$ as follows:

$$\sigma_{zx}^x = \frac{-a+c}{2}\sin 2\theta,$$
$$\sigma_{zx}^y = \frac{1}{2}(a+c-2d) - \frac{1}{2}(a-c)\cos 2\theta,$$
$$\sigma_{zx}^z = -\frac{a+b+c}{\sqrt{2}}\cos\theta. \quad (6)$$

Figures 3(b-d) show the respective variations of $\sigma_{zx}^k$ as functions of $\theta$ for GaNMn$_3$ (110). Similar angular dependences of $\sigma_{zx}^k$ for the ground states of NiNMn$_3$, SnNMn$_3$, and PtNMn$_3$ (110) films can be found in Supplemental Material [50].

The predicted efficient generation of the longitudinal and transverse currents with sizable spin polarization allows promising spintronic devices based on noncollinear antiferromagnetic $A$NMn$_3$. Here we propose two types of spintronic devices, as shown in Figure 4. The first one is an antiferromagnetic tunnel junction [38], where the two $A$NMn$_3$ electrodes are separated by an insulating nonmagnetic layer [Fig. 4(a)]. The CPP longitudinal spin polarized current is controlled by the relative orientation of the magnetic order parameters in the $A$NMn$_3$ reference and free layers. Due to a large spin polarization of the longitudinal current, the TMR effect is expected to be sizable and hence can be used to efficiently detect the magnetic order parameter in $A$NMn$_3$. The spin polarized current can be also used to generate the spin-transfer torque for switching $A$NMn$_3$ [70, 77-79].

The second spintronic device is a CIP spin-torque device, where an ANMn$_3$ (110) layer is used as a spin source to generate an out-of-plane spin current which enters the top ferromagnetic layer and exerts a torque on its magnetization (Fig. 4(b)). Since $A$NMn$_3$ (110) layer exhibits large $\Phi(\sigma_{zx}^z)$, the spin current can carry sizable spin component collinear to the spin current direction, which is necessary for switching a ferromagnet with perpendicular anisotropy required for high-density spintronic devices.

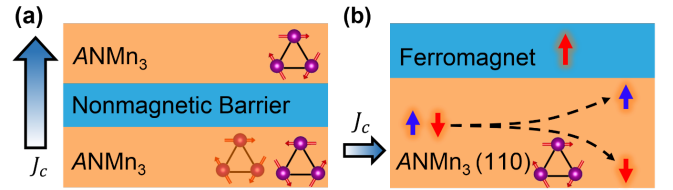

**FIG. 4. (a)** The schematic of an antiferromagnetic tunnel junction using $A$NMn$_3$ as the electrodes, where the transport of the out-of-plane longitudinal spin polarized current is controlled by the relative orientation of the magnetic moments between two $A$NMn$_3$ layers. **(b)** The schematic of a spin torque device with CIP geometry. An in-plane charge current passes through $A$NMn$_3$ (110) layer and generates an out-of-plane spin current carrying sizable spin polarization collinear to the spin current direction. This spin current can exert a torque on the perpendicular magnetization in the top ferromagnetic layer for an efficient switching.

We note that the symmetries of bulk antiperovskites are lowered at the interfaces of such devices which may affect spin-dependent transport properties, such as the TMR and spin-transfer torque. It was found, however, that in realistic antiferromagnet/nonmagnet systems, these effects were not essential for the device performance [11, 80, 81]. We expect therefore that the predicted bulk properties would largely control functional properties of the device structures based on antiferromagnetic antiperovskites.

In conclusion, based on first-principles density functional theory calculations, we have predicted that the noncollinear antiferromagnetic antiperovskites $A$NMn$_3$ ($A$ = Ga, Ni, Sn, and Pt) support electric currents with sizable spin polarization. We found that the calculated spin polarization of the longitudinal currents can be comparable to that in widely used ferromagnetic metals, which makes the antiperovskites promising for using in antiferromagnetic tunnel junction and spin transfer torque devices. Furthermore, we demonstrated that by controlling the film growth direction, the out-of-plane transverse spin currents with sizable charge-to-spin conversion efficiencies can be achieved, which implies that the $A$NMn$_3$ compounds can serve as effective spin source materials. These properties make noncollinear antiferromagnetic antiperovskites promising for realistic applications in spintronics.

**Acknowledgements:** This work was supported by the Office of Naval Research (ONR grant N00014-20-1-2844) and by the National Science Foundation (NSF) through EPSCoR RII Track-1 (NSF Award OIA-2044049) program. Computations were performed at the University of Nebraska Holland Computing Center. The figures were created using VESTA [82], FermiSurfer [83] and Matplotlib.

\* dfshao@unl.edu
† tsymbal@unl.edu